\documentclass[pra,superscriptaddress,showpacs,twocolumn,aps,10pt]{revtex4-1}

\usepackage{amsfonts,amssymb,amsmath}
\usepackage{graphics,graphicx,epsfig}
\usepackage{amsthm}

\def\identity{\leavevmode\hbox{\small1\kern-3.8pt\normalsize1}}

\newcommand{\be}{\begin{eqnarray}}
\newcommand{\ee}{\end{eqnarray}}

\newcommand{\ket}[1]{\left | #1 \right\rangle}
\newcommand{\bra}[1]{\left \langle #1 \right |}

\newcommand{\Tr}{\mathrm{Tr}}

\newcommand{\proj}[1]{\ket{#1}\bra{#1}}
\renewcommand{\epsilon}{\varepsilon}

\usepackage{graphicx}
\begin{document}

\title{Entanglement witnesses with variable number of local measurements}

\author{Wies\l aw~Laskowski} 
\affiliation{Institute of Theoretical Physics and Astrophysics, University of Gda\'nsk, 80-952 Gda\'nsk, Poland}

\author{Marcin Markiewicz} 
\affiliation{Institute of Theoretical Physics and Astrophysics, University of Gda\'nsk, 80-952 Gda\'nsk, Poland}

\author{Tomasz Paterek}
\affiliation{School of Physical and Mathematical Sciences, Nanyang Technological University, Singapore}
\affiliation{Centre for Quantum Technologies, National University of Singapore, Singapore}

\author{Ryszard Weinar}
\affiliation{Institute of Theoretical Physics and Astrophysics, University of Gda\'nsk, 80-952 Gda\'nsk, Poland}

\begin{abstract}
We present a class of entanglement identifiers which has the following experimentally friendly feature:
once the expectation value of the identifier exceeds some definite limit, we can conclude the state is entangled,
even if not all measurements defining the identifier have been performed.
These identifiers are in the form of sums of nonnegative functions of correlations in a quantum state, mostly squares of correlations, and we illustrate their use and strengths on various examples.
\end{abstract}


\maketitle

\section{Introduction}

Entanglement witness is traditionally an observable with expectation value below certain bound for all separable states and exceeding the bound for some entangled states~\cite{WITNESS,HORODECCY}.
Every entangled state can be detected in this way by a suitably designed entanglement witness.
Furthermore, the witness operator can always be expressed in terms of local measurements and therefore entanglement can be certified if the corresponding linear combination of suitable correlations and averages of local measurements exceeds the bound set by separable states.
Note however that usually all the correlations defining the witness have to be measured.
It may happen that the bound is violated already for a subset of local measurements defining the witness but together with its other measurements the separability bound is not exceeded.

For this reason the entanglement identifiers that sum nonnegative functions of correlations in a quantum state are relevant from practical perspective.
As soon as their separability bound is exceeded, even for a subset of measurements defining the identifier, entanglement of the state is confirmed as any future measurements can only add to the present value of the identifier.

Here we utilise the general results on entanglement detection presented in Ref.~\cite{FRIENDLY}
in order to derive a set of local measurements revealing entanglement of a state given its closest separable state according to Hilbert-Schmidt distance.
This approach is equivalent to the traditional entanglement witnesses and next we propose a procedure to obtain a quadratic criterion from the linear witness.
This criterion is different from the original linear one and in general detects a different set of entangled states.
Since finding the closest separable state is known to be a hard problem even in a bipartite scenario~\cite{SEP_NP},
we also discuss a simpler task in which the closest separable state is replaced by the closest state with positive partial transposition (PPT)~\cite{PPT}.
Though it should be noted that entangled PPT states exist which are still very distant from the set of separable states \cite{BS2010}.
We utilise the algorithm for finding the closest PPT states~\cite{VDM} to establish combinations of correlations revealing entanglement in NPT states, i.e. states with negative partial transposition.
All this is illustrated with examples, some of which go beyond the proposed procedure and sum other positive functions of correlations  as well as detect entanglement without knowledge of the closest separable state.

\section{Motivation}

We begin with an explicit example of a linear entanglement witness that illustrates the need for making all measurements in its definition.
It is sufficient to consider states of two qubits (two-level quantum systems).
All such states can be represented in terms of expectation values of local Pauli operators as follows:
\begin{equation}
\rho = \frac{1}{4} \sum_{\mu, \nu = 0}^3 T_{\mu \nu} \sigma_\mu \otimes \sigma_\nu,
\end{equation}
where $\sigma_{\mu} \in \{\openone, \sigma_x, \sigma_y, \sigma_z \}$ and $T_{\mu \nu} = \Tr[\rho(\sigma_\mu \otimes \sigma_\nu)]$ is directly experimentally accessible correlation function.

Consider the following linear combination of correlation functions and local averages:
\begin{equation}
T_{xx} + T_{zz} - T_{z0} + T_{0z} \le \frac{3}{2},
\label{LINEAR_WIT}
\end{equation}
where the bound holds for separable two-qubit states, i.e. states that can be represented as $\rho_{\mathrm{sep}} = \sum_i p_i \rho_i^A \otimes \rho_i^B$ with $p_i$'s being probabilities.
Eq. (\ref{LINEAR_WIT}) is an entanglement witness that, for example, detects the Bell state $\ket{\phi^+} = \frac{1}{\sqrt{2}}(\ket{00} + \ket{11})$, where $\ket{0}$ ($\ket{1}$) denotes the eigenstate of $\sigma_z$ local Pauli operator with eigenvalue $+1$ ($-1$).
The witness is constructed in such a way that the product state $\ket{11}$ violates the bound of $\frac{3}{2}$ if we measure correlations $T_{zz}=1$ and $T_{z0}=-1$ only, but of course the bound is satisfied if the remaining correlations are taken into account.
In this work we present  entanglement identifiers such that once the separability bound is exceeded for a subset of measurements defining the identifier the other measurements can be skipped.

\section{Entanglement criteria}

For simplicity of presentation we shall demonstrate the results using states of qubits, but the techniques are applicable to any systems.
Arbitrary quantum state is represented by a density operator $\rho$ which can be regarded as an element of a Hilbert-Schmidt vector space with trace inner product.
We call a \emph{standard basis} in this space a basis composed of vectors $| e_{\mu} ) \equiv \ket{m} \bra{n}$ for individual qubits, where $m,n=0,1$ and $\mu= 2m+n$,
i.e. the rounded kets denote vectors in the Hilbert-Schmidt space of mixed states whereas the usual kets denote vectors in the Hilbert space of pure single qubit states.

The elements of the standard basis in general do not admit spectral decomposition and therefore they do not correspond to directly measurable quantities.
From experimental perspective another basis plays a crucial role: $| f_\mu ) \in \{\openone, \sigma_x, \sigma_y, \sigma_z\}$,
which is composed of Pauli operators.
Note that the vectors of this basis are not normalized, instead we have $(f_{\mu} | f_\nu) = \Tr(\sigma_\mu \sigma_\nu) =  2 \delta_{\mu \nu}$.
These correspond to physical observables and since their averages uniquely define a quantum state of a qubit
we shall call this basis a \emph{tomographic basis}.

Let us briefly describe the main result of Ref. \cite{FRIENDLY} in a slightly more general context than original.
If a set $\mathcal{S}$ of quantum states is convex and compact then
the following necessary and sufficient condition verifies whether $\rho$ belongs to $\mathcal{S}$:
\begin{equation}
\max_{\rho_s \in \mathcal{S}} (\rho | G | \rho_s) < (\rho | G | \rho) \quad \mathrm{ iff } \quad \rho \notin \mathcal{S},
\label{RHO_CRIT}
\end{equation}
where $G$ is a positive semidefinite metric acting in the Hilbert-Schmidt space of mixed states.
In Ref. \cite{FRIENDLY} this condition has been proved for the convex set of separable quantum states,
but the proof for any other convex set goes along the same lines.

Here we focus on a specific choice of metric $G$ that reduces criterion (\ref{RHO_CRIT}) to entanglement witnesses.
This happens for $G$ being a projector in the Hilbert-Schmidt space, $G = | \delta )(\delta|$, 
where $\delta = \rho - \rho_0$ is the traceless operator with $\rho_0$ being the closest state in $\mathcal{S}$ to $\rho$.
It was shown that  $(\rho | \delta) > 0$, and therefore this choice of metric transforms condition (\ref{RHO_CRIT}) into:
\begin{equation}
\max_{\rho_s \in \mathcal{S}} (\rho - \rho_0 | \rho_s) < (\rho - \rho_0 | \rho) \quad \mathrm{ iff } \quad \rho \notin \mathcal{S}.
\label{PROJ_CRIT}
\end{equation}
In this form it is more transparent that the condition is necessary and sufficient.
If $\rho$ belongs to $\mathcal{S}$ then the difference $\rho - \rho_0$ vanishes and both sides of the criterion are equal to zero.
If $\rho$ does not belong to $\mathcal{S}$ we briefly repeat the reasoning of Ref. \cite{FRIENDLY} for completeness.
In this case the assumption implies positivity of the length $||\delta||^2 > 0$, i.e. $(\delta | \rho_0) < (\delta | \rho)$.
To derive (\ref{PROJ_CRIT}) we have to show that $(\delta | \rho_s) \le (\delta | \rho_0)$.
To this end, define the state $\rho_\lambda = (1-\lambda) \rho_0 + \lambda \rho_s$.
Since $\rho_0$ is the closest state, we have $||\rho - \rho_\lambda||^2 \ge ||\delta||^2$ with equality only for $\lambda = 0$.
Therefore we require $\frac{\partial}{\partial \lambda} ||\rho - \rho_\lambda ||^2 \ge 0$ at $\lambda = 0$.
This is equivalent to the condition $(\delta | \rho_s) \le (\delta | \rho_0)$.

\subsection{Measurements revealing entanglement}

Suppose that the closest separable state, $\rho_0$, to the state $\rho$ is known.
In order to find measurements which confirm entanglement we change the basis in which density operators are written from standard to tomographic.
For example, any state of $N$ qubits can be represented in the tensor product of the tomographic bases as
\begin{equation}
| \rho ) = \frac{1}{2^N} \sum_{\mu_1, \dots, \mu_N = 0}^3 T_{\mu_1 \dots \mu_N} |f_{\mu_1} \dots f_{\mu_N}),
\end{equation}
where $|f_{\mu_1} \dots f_{\mu_N})$ shortly denotes $|f_{\mu_1}) \otimes \dots \otimes  |f_{\mu_N})$
and the components are directly experimentally accessible correlation functions and 
averages of local Pauli measurements $T_{\mu_1 \dots \mu_N} = (f_{\mu_1} \dots f_{\mu_N} | \rho) = \Tr(\rho \sigma_{\mu_1} \otimes \dots \otimes \sigma_{\mu_N})$.
A set of $4^N$ elements $T_{\mu_1 \dots \mu_N}$ forms the so-called extended correlation tensor $\hat T$, which is just an alternative representation of a quantum state $\rho$.

Any operator can be decomposed in the tomographic basis.
In particular, let us denote by $\hat D$ the extended correlation tensor of the traceless operator $\delta = \rho - \rho_0$.
In this language, the trace inner products of the entanglement criteria become scalar products between correlation tensors.
For example, in criterion (\ref{PROJ_CRIT}) for $N$ qubits the products are of the form
\begin{equation}
(\rho - \rho_0 | \rho) = \frac{1}{2^N} \sum_{\mu_1, \dots, \mu_N = 0}^3 D_{\mu_1 \dots \mu_N} T_{\mu_1 \dots \mu_N}.
\end{equation}
Since tensor $\hat D$ is present on both sides of the entanglement criterion,
its particular numerical values are not essential, only the relative weights matter for the detection.
Another important message is that if $D_{\mu_1 \dots \mu_N}$ vanishes for some indices,
the corresponding correlations are not relevant for entanglement detection.
Such measurements can be skipped in an experiment verifying entanglement of state $\rho$.

\subsection{Quadratic identifiers}

In the previous section we showed how to construct a linear entanglement witness from the knowledge of the closest separable state.
In order to construct a quadratic entanglement identifier we propose the following procedure.
Take the linear witness of Eq. (\ref{RHO_CRIT}) with $G = | \delta )(\delta|$.
When written in the tomographic basis this metric has matrix elements:
\begin{equation}
(f_{\mu_1} \dots f_{\mu_N} | G | f_{\mu_1'} \dots f_{\mu_N'}) = D_{\mu_1 \dots \mu_N} D_{\mu_1' \dots \mu_N'}.
\end{equation}
We now erase all off-diagonal elements of this metric and define a new metric, $H$, being the diagonal part of the previous one:
\begin{equation}
(f_{\mu_1} \dots f_{\mu_N} | H | f_{\mu_1'} \dots f_{\mu_N'}) \equiv D_{\mu_1 \dots \mu_N}^2 \delta_{\mu_1 \mu_1'} \dots \delta_{\mu_N \mu_N'}.
\end{equation}
Finally, we optimise the left-hand side of condition (\ref{RHO_CRIT}) with metric $H$:
\begin{eqnarray}
& \max_{\hat T^\mathrm{sep}} & \sum_{\mu_1, \dots, \mu_N = 0}^3 T_{\mu_1 \dots \mu_N} D_{\mu_1 \dots \mu_N}^2 T_{\mu_1 \dots \mu_N}^{\mathrm{sep}} \nonumber \\
& < & \sum_{\mu_1, \dots, \mu_N = 0}^3 T_{\mu_1 \dots \mu_N}^2 D_{\mu_1 \dots \mu_N}^2,
\label{WITH_H}
\end{eqnarray}
in order to check if indeed entanglement can be detected.

\section{Examples}

Let us illustrate the methods discussed here with a few concrete examples.
We also utilize the algorithm by Verstraete \emph{et al.}~\cite{VDM} to find the closest PPT or separable states in some new cases.

\subsection{Werner state}

Consider the Werner state of two qubits:
\begin{equation}
\rho_{\mathrm{Werner}} = p \ket{\phi^+} \bra{\phi^+} + (1-p) \frac{1}{4} \openone_4,
\label{WERNER}
\end{equation}
where $\ket{\phi^+} = \frac{1}{\sqrt{2}}(\ket{00} + \ket{11})$ is the maximally entangled Bell state and $\openone_4$ denotes identity matrix for two qubits.
Its closest separable state is known to be~\cite{WT99}:
\begin{equation}
\rho_0 =
\left(
\begin{array}{cccc}
\frac{1}{3} & 0 & 0 & \frac{1}{6} \\
0 & \frac{1}{6} & 0 & 0 \\
0 & 0 & \frac{1}{6} & 0 \\
\frac{1}{6} & 0 & 0 & \frac{1}{3}
\end{array}
\right),
\quad \textrm{ for } \quad p > \frac{1}{3},
\end{equation}
and $\rho_0 = \rho$ for $p \le \frac{1}{3}$.
Note that the $\rho_0$ state is independent of $p$ as soon as $p > \frac{1}{3}$.
This is intuitively understood as follows:
the Werner states for different values of $p$ lie on a line between completely mixed state and the maximally entangled state.

Calculation of tensor $\hat D$, for the operator $\rho - \rho_0$,
reveals that only three measurements are required to prove entanglement for the whole family of Werner states.
Namely, all the correlations $D_{\mu_1 \mu_2}$ vanish except for $D_{xx} = - D_{yy} = D_{zz} = p - \frac{1}{3}$.

In order to arrive at the linear witness we explicitly calculate the maximum over separable states in criterion (\ref{PROJ_CRIT}).
Since the scalar product is linear, it is sufficient to calculate the maximum over pure product states.
Further, since $\hat D$ is present on both sides of (\ref{PROJ_CRIT}) and $p - \frac{1}{3}$ is positive for $p>\frac{1}{3}$,
we divide both sides by this number and calculate the maximum of the left-hand side to be
$T_{x_1} T_{x_2} - T_{y_1} T_{y_2} + T_{z_1} T_{z_2} \le 1$,
where the maximum follows once we write components of local tensors (Bloch vectors) in spherical coordinates,
$(T_x,T_y,T_z) = (\sin \theta \cos \varphi, \sin \theta \sin \varphi, \cos \theta)$.
Finally, the Werner states (\ref{WERNER}) are entangled if and only if
\begin{equation}
T_{xx} - T_{yy} + T_{zz} > 1.
\label{WERNER_CRIT}
\end{equation}

The quadratic identifier also detects all the states of this family.
Since all the elements of the diagonal metric $H$ are the same and present on both sides of the condition (\ref{WITH_H}) they cancel out.
The left-hand side therefore reads $L = T_{xx} T_{x_1} T_{x_2} + T_{yy} T_{y_1} T_{y_2} + T_{zz} T_{z_1} T_{z_2}$ and its maximum for the Werner state is $p$.
On the right-hand side we have the sum of squared correlations.
Therefore, the state is entangled if
\begin{equation}
T_{xx}^2 + T_{yy}^2 + T_{zz}^2 > p,
\label{QUAD_WERNER}
\end{equation}
which holds exactly for $p > \frac{1}{3}$.
Note that for many Werner states only two measurements are sufficient to reveal entanglement.

\subsection{Bell diagonal states}

Consider any two-qubit state with completely mixed states of individual qubits.
Such states are diagonal in the Bell basis:
\begin{equation}
\rho_{\mathrm{BD}} = \sum_{j=1}^4 p_j \proj{\psi_j},
\end{equation}
where $p_j$'s are probabilities and $\ket{\psi_j}$'s describe the Bell basis, e.g. $\ket{\psi_1} = \frac{1}{\sqrt{2}}(\ket{00} + \ket{11})$,
$\ket{\psi_2} = \frac{1}{\sqrt{2}}(\ket{00} - \ket{11})$, $\ket{\psi_3} = \frac{1}{\sqrt{2}}(\ket{01} + \ket{10})$, and $\ket{\psi_4} = \frac{1}{\sqrt{2}}(\ket{01} - \ket{10})$.
The closest separable state to $\rho_{\mathrm{BD}}$ has been found in Ref.~\cite{WT99}
but for our purposes it is better to present the difference:
\begin{equation}
\rho_{\mathrm{BD}} - \rho_0 = \frac{2}{3} p_j (\proj{\psi_j} - \tfrac{1}{4} \openone_4), \textrm{ for } j = 1,2,3,4.
\label{BD_CLOSEST_DIFF}
\end{equation}
It turns out that there are four ways of expressing the closest separable state depending on which particular Bell state we would like to use for description.
These four ways lead to four different tensors $\hat D_j$ given by:
\begin{equation}
\hat D_j = \frac{2}{3} p_j \hat T_j, \textrm{ for } j = 1,2,3,4,
\end{equation}
where $\hat T_j$ denotes correlation tensor of the Bell state $\ket{\psi_j}$, which in this case is a $3 \times 3$ matrix with elements given by $T_{kl}$ for $k,l = x,y,z$.
Since the positive factor $\frac{2}{3} p_j$ appears on both sides of criterion (\ref{PROJ_CRIT}) it cancels out and it is easy to verify that the maximum of the left-hand side is then given by the unity.
Finally, a Bell diagonal state is entangled if any of the following inequalities is satisfied:
\begin{eqnarray}
- T_{xx} + T_{yy} + T_{zz} & > & 1, \nonumber \\
T_{xx} - T_{yy} + T_{zz} & > & 1, \nonumber \\
T_{xx} + T_{yy} - T_{zz} & > & 1, \nonumber \\
- T_{xx} - T_{yy} - T_{zz} & > & 1.
\label{DB_SET_INEQ}
\end{eqnarray}

This example leads to an interesting criterion in which nonnegative functions of correlations are being summed, but they are not the squares of the correlations.
Namely we will now show that the four inequalities above are equivalent to the Horodeckis necessary and sufficient condition for entanglement of Bell diagonal states~\cite{BD_HOR1996}:
\begin{equation}
|T_{xx}| + |T_{yy}| + |T_{zz}| > 1.
\label{BD_HOR}
\end{equation}
Note that in order to introduce the absolute values of the correlations
the set of inequalities (\ref{DB_SET_INEQ}) has to be augmented with the four inequalities
obtained by multiplying only the left-hand side of each inequality in (\ref{DB_SET_INEQ}) by $-1$.
We demonstrate that for physical states these four new inequalities are never satisfied and therefore entanglement is fully characterised by (\ref{DB_SET_INEQ}) or equivalently (\ref{BD_HOR}).

For concreteness consider the inequality obtained from the last expression in  (\ref{DB_SET_INEQ}), i.e. $T_{xx} + T_{yy} + T_{zz} > 1$.
Argument along the same lines applies to any other inequality.
Assume there exists a state $\eta$ for which this inequality is satisfied and calculate the probabilities to observe different measurement outcomes when $\eta$ is measured by a Bell state analyser.
In particular probability to observe the result corresponding to the singlet state reads $\frac{1}{4}(1-T_{xx} - T_{yy} - T_{zz})$ and it would be negative for the hypothetical state $\eta$. Thus, for all physical states $T_{xx} + T_{yy} + T_{zz} \le 1$.

\subsection{Isotropic states}

This example is similar to the Werner states in that there is only one state $\rho_0$ closest to all the states of the family.
This is a one parameter family of states for two $d$-level systems called isotropic states~\cite{ISO1,ISO2}:
\begin{equation}
\rho_{\mathrm{Iso}} = p \proj{\psi^+_d} + (1-p) \frac{1}{d^2} \openone_{d^2},
\end{equation}
where $\frac{1}{d^2} \openone$ describes completely mixed state of two qudits and $\ket{\psi^+_d} = \frac{1}{\sqrt{d}}\sum_{j=0}^{d-1} \ket{jj}$ is a maximally entangled state.
The closest separable state is found to be of the same form having $p = \frac{1}{d+1}$~\cite{BDHK2005}.

We represent a state of two $d$-level systems in the tomographic basis composed of generalised Pauli operators $\lambda_j$ with $j=1,\dots,d^2$.
$\lambda_1$ is chosen as identity and all others as traceless hermitian observables that satisfy $\Tr(\lambda_i \lambda_j) = 2 \delta_{ij}$ 
and have entries that are either purely real or purely imaginary.
For $d=3$ one may think about Gell-Mann matrices and for any $d$ about their generalisations.
The correlation tensor elements are the expansion coefficients of a state in this tomographic basis:
\begin{equation}
T_{ij} = \frac{d}{2(d-1)}\mathrm{Tr}( \rho \lambda_i\otimes\lambda_j),
\end{equation}
where the factor of $\frac{d}{2(d-1)}$ is introduced in order to keep the correlations in the range $| T_{ij} | \le 1$.

The state $\rho_{\mathrm{Iso}}$ has diagonal correlation tensor, which apart from the element $T_{00}$, can be rewritten as:
\begin{equation}
T_{ij}=\frac{pd}{2(d-1)}\langle\psi^+_d|\lambda_i\otimes\lambda_j|\psi^+_d\rangle=\pm\frac{p}{d-1}\delta_{ij}.
\end{equation}
Note that precise location of plus and minus signs cannot be inferred from this formula and indeed both signs are compatible with the definition of $\lambda_j$'s given above.
A linear witness can be constructed if additional constraints on the structure of  $\lambda_j$'s are assumed.
For example, taking the matrices of Ref.~\cite{EULER_SUN} fixes the positions of pluses and minuses and allows explicit construction of the linear witness.
In contradistinction, we offer here quadratic identifier that detects entanglement of the isotropic states
in the full range without any further restrictions on $\lambda_j$'s.

Following our procedure, tensor $\hat D$ has the following non-vanishing components, $D_{ii} = \pm (\frac{p}{d-1} - \frac{1}{d^2-1})$, which squared appear on both sides of condition (\ref{WITH_H}) and cancel out.
Maximization of the left-hand side gives the biggest correlation in the state, $T_{\max}$, and the quadratic identifier simplifies to
\be
\sum_{j=1}^{d^2-1}T_{jj}^2>T_{\max},
\label{QUAD_ISO}
\ee
where the unknown sign of correlations is no longer a problem. 
For the isotropic states $T_{\max} = \frac{p}{d-1}$, and entanglement is detected for $p>\frac{1}{d+1}$, which is exactly the same as in~\cite{BDHK2005}.

\subsection{Bound entanglement}

To show the strength of the criterion that sums squares of correlations we now apply it to detect bound entanglement.
Consider the PPT states proposed by Horodecki \cite{HORODECKISTATE1} and generalized in \cite{HORODECKISTATE2}.
For $C^3\otimes C^3$ (two qutrits) the state can be written as
\be
\rho_a =\frac{1}{8a+1}
\left(
\begin{array}{ccc|ccc|ccc}
a &  \cdot & \cdot  & \cdot  & a &  \cdot &  \cdot & \cdot  & a \\ 
 \cdot & a & \cdot  & \cdot  & \cdot  &  \cdot & \cdot  &  \cdot &  \cdot \\
 \cdot &  \cdot & a & \cdot  & \cdot  & \cdot  & \cdot  &  \cdot &  \cdot \\\hline
 \cdot & \cdot  & \cdot  & a &  \cdot & \cdot  &  \cdot &  \cdot &  \cdot \\
a &  \cdot &  \cdot & \cdot  & a &  \cdot &  \cdot &  \cdot &  a \\
 \cdot & \cdot  &\cdot   & \cdot  &  \cdot & a &   \cdot&  \cdot &  \cdot \\ \hline
 \cdot & \cdot  & \cdot  & \cdot  &  \cdot & \cdot  & b &  \cdot &  c \\
 \cdot & \cdot  & \cdot  &  \cdot &  \cdot & \cdot  &  \cdot & a &  \cdot \\
a & \cdot  & \cdot  &  \cdot & a & \cdot  & c & \cdot  &  b \\

\end{array}
\right),
\label{PPT_STATE}
\ee 
where $b=\frac{1+a}{2}$, $c=\frac{\sqrt{1-a^2}}{2}$ and $a\in [0,1]$. 
The state is PPT in the whole range of $a$, it is separable for $a = 0$ and $a=1$, and entangled for $a\in(0,1)$ as shown in Ref.~\cite{HORODECKISTATE1}.

By taking the Gell-Mann matrices as tomographic basis for qutrits,
the correlation tensor elements defined by $T_{ij}=\frac{3}{4} \mathrm{Tr}(\rho_a\lambda_i\otimes\lambda_j)$ can be easily determined. 
The non-vanishing elements are
\be \nonumber
T_{11}&=&T_{22}=T_{33}=-T_{44}=-T_{55}=-T_{66}=\frac{3a}{2+16a}, \\ \nonumber
T_{82}&=&-\frac{\sqrt{3(1-a)}}{2+16a}, ~ T_{87}=-\frac{\sqrt{3}(1-a)}{4+32a}, ~ T_{88}=\frac{1-a}{4+32a}.
\ee

\begin{figure}[!tp]
\centering
\includegraphics[scale=0.55]{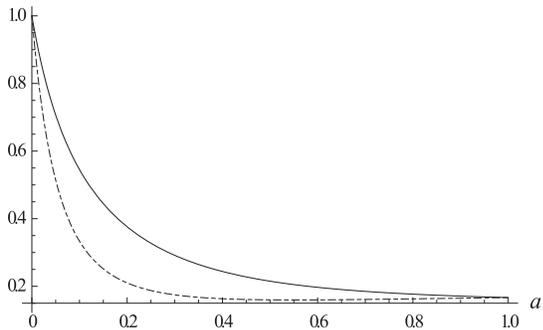}
\caption{Detection of bound entanglement. The upper solid line presents the sum of $T_{ij}^2$ for the state (\ref{PPT_STATE}) as a function of state parameter $a$.
The lower dashed curve is the upper bound on maximal correlation of this state.
According to criterion (\ref{CRIT_NORM_T}) we detect entanglement in the whole range of $a \in (0,1)$.}
\label{figppt}
\end{figure}

As entanglement criterion consider the following generalisation of (\ref{QUAD_WERNER}) and (\ref{QUAD_ISO}):
\begin{equation}
\sum_{i,j=1}^{d^2-1} T_{ij}^2 > T_{\max},
\label{CRIT_NORM_T}
\end{equation}
where $T_{\max}$ is the biggest correlation in the state, i.e. the biggest overlap between the correlation tensor of the state and the correlation tensor of a product state, $T_{\max} = \max_{T^{\mathrm{prod}}} \sum_{i,j=1}^{d^2-1} T_{ij} T^{\mathrm{prod}}_{ij}$.
If (\ref{CRIT_NORM_T}) holds the state cannot be separable.
Fig.~\ref{figppt} shows that the sum of squared correlation tensor components of $\rho_a$ is greater than the maximal singular value of matrix composed of $T_{ij}$ elements, for all $a \in (0,1)$.
Note that the maximal singular value might be greater than the actual $T_{\max}$ because the tensor product of vectors for which it is achieved might be unphysical.
Surprisingly, this in principle weaker condition detects all entangled states $\rho_a$.

This example also shows that we do not always have to look for the nearest separable state to build appropriate metric.
 
\subsection{NPT witness}

Since the set of PPT states is convex and compact, the criterion (\ref{PROJ_CRIT}) can be interpreted as an NPT witness.
Moreover, if the dimension of the total system is not greater than six,
the sets of separable states and PPT states coincide~\cite{WITNESS}.

A practical tool to reveal measurements showing entanglement in NPT states
consists of finding the closest PPT state and then changing the basis to the tomographic one.
The closest PPT state with respect to the Hilbert-Schmidt distance can be found using the algorithm of Verstraete \emph{et al.} \cite{VDM}.
The algorithm returns the closest partially transposed matrix,
which almost always is positive semidefinite (if the dimension is not too high) and therefore a valid PPT state.

The algorithm consists of the following steps \cite{VDM}:
\begin{itemize}

\item[(i)] Calculate the eigenvalue decomposition of the partially transposed state $\rho$: $\rho^{\mathrm{PT}} = V \mathcal{D} V^{\dagger}$,
where $\mathcal{D}$ is a diagonal matrix with entries $d_i$.
\item[(ii)] Define diagonal positive semidefinite matrix $\mathcal{D}'$ with entries $d_i' = d_i + \lambda$ if $d_i$ is non-negative, and $d_i' = 0$ if $d_i$ is negative.

\item[(iii)] Calculate $\lambda$ from normalization $\Tr(\mathcal{D}') = 1$.

\item[(iv)] The closest partially transposed matrix to $\rho$ is given by $\rho_0 = (V \mathcal{D}' V^{\dagger})^{\mathrm{PT}}$.

\end{itemize}
If such defined $\rho_0$ is positive semidefinite, it is the closest PPT state.

\subsubsection{Colored noise}

We apply this method to reveal entanglement of a pure entangled state mixed with colored noise:
\begin{equation}
\rho_{\mathrm{Color}} = p \proj{\phi^+} + (1-p) \proj{01}.
\label{COLOR}
\end{equation}
To illustrate the steps we fix $p=\frac{2}{3}$ and next discuss numerical investigations in the whole range of $p$.
For $p=\frac{2}{3}$ the closest separable state found using the above mentioned algorithm reads:
\begin{equation}
\rho_{0} = \frac{1}{9}\left(\begin{array}{cccc}
 \frac{7 - \sqrt{5}}{2} & 0 & 0 &  \frac{5 + 2\sqrt{5}}{5} \\
0&   \frac{15 + 7 \sqrt{5}}{10}& 0 & 0 \\
0& 0& \frac{5+3\sqrt{5}}{10} & 0 \\
  \frac{5 + 2\sqrt{5}}{5} &  0& 0&  \frac{7 - \sqrt{5}}{2}
\end{array}\right).
\end{equation}
The corresponding tensor $\hat D$ has the following non-vanishing components:
\begin{eqnarray}
-D_{xx} & = & D_{yy} = \frac{4}{45}(-5+\sqrt{5}), \nonumber \\
D_{zz} & = & \frac{2}{9}(\sqrt{5}-1), \nonumber \\
D_{0z} & = & - D_{z0} = \frac{2}{45}(-5+\sqrt{5}).
\end{eqnarray}
The maximal value of the left-hand side of condition (\ref{PROJ_CRIT}) is $\frac{2}{9} (\sqrt{5}-1)$ and the linear witness takes the form:
\begin{equation}
2 T_{xx} - 2 T_{yy} + \sqrt{5} T_{zz} + T_{z0} - T_{0z} > \sqrt{5}.
\end{equation}
It is easy to check that it indeed reveals entanglement of state (\ref{COLOR}) with $p=\frac{2}{3}$.
Note that additionally to correlations this witness takes advantage of non-vanishing local averages for $\sigma_z$ Pauli measurements.
Entanglement of the state (\ref{COLOR}) is also revealed with the corresponding quadratic identifier:
\begin{equation}
4 T_{xx}^2 + 4 T_{yy}^2 + 5 T_{zz}^2 + T_{z0}^2 + T_{0z}^2 > \frac{35}{13}.
\end{equation}
We have also verified numerically that the linear witness detects entanglement of all the states of the family (\ref{COLOR}), i.e. for all positive $p$.
The corresponding quadratic identifiers are also quite powerful and detect entanglement of all these states for $p > \frac{1}{2}$.

\subsection{W state}

Finally, we give a multipartite example.
Consider the three qubit W state~\cite{W}: $$|W\rangle =\frac{1}{\sqrt{3}}(|100\rangle+|010\rangle+|001\rangle).$$
Its closest separable state is given by~\cite{XXX}:
 $$\rho_{0} = \frac{23}{63} |W\rangle \langle W| + \frac{40}{63} \cdot \frac{1}{8} \openone.$$
In this case tensor $\hat D$ has the following non-vanishing components: 
$D_{003} = D_{030} = D_{300} = \frac{40}{189}$, $D_{033} = D_{303} = D_{330} = -\frac{40}{189}$, $D_{011} = D_{110} = D_{101} = D_{022}  = D_{202} = D_{220} =D_{113} =D_{131}=D_{311}=D_{223}=D_{232}=D_{322} = \frac{80}{189}$, and $D_{333}=-\frac{40}{63}$.
Linear condition takes the form:
\begin{eqnarray}
&&T_{003} + T_{030} + T_{300}  + 2 T_{101} + 2 T_{110} + 2 T_{011} \nonumber\\
&&+ 2 T_{022} + 2 T_{202} + 2 T_{220} - T_{033} - T_{303}- T_{330}\nonumber\\
&&+ 2 T_{113} + 2 T_{131} +  2 T_{311} + 2 T_{223} + 2 T_{232} \nonumber\\  
&&+ 2 T_{322}  - 3 T_{333} > \frac{23}{3}, 
\end{eqnarray}
which gives $21> \frac{23}{3}$ and proves entanglement. 

Quadratic condition is given by:
\begin{eqnarray}
&&T_{003}^2 + T_{030}^2 + T_{300}^2 + 4 T_{101}^2 + 4 T_{110}^2 + 4 T_{011}^2 \nonumber \\
&&+ 4 T_{022}^2 + 4 T_{202}^2 + 4 T_{220}^2  + T_{033}^2 + T_{303}^2+ T_{330}^2\nonumber \\
&&+ 4 T_{113}^2 + 4 T_{131}^2 +  4 T_{311}^2 + 4 T_{223}^2 + 4 T_{232}^2   \nonumber \\
&&+ 4 T_{322}^2  +9 T_{333}^2 > \frac{29}{3}, 
\end{eqnarray}
and it also detects entanglement, since in this case: $31 > \frac{29}{3}$.

\section{Conclusions}

We presented examples of entanglement identifiers which  sum only nonnegative functions of correlations in a quantum state.
In this way, as soon as the bound of the identifier set by separable states is exceeded, one can conclude the state is entangled without a need for any further measurements.
These identifiers are quite powerful and easily detect bound entangled states.

\acknowledgments

This work is dedicated to Professor Marek \.Zukowski
on the occasion of his 60th birthday.
We thank Otfried G\"uhne for helpful discussions.
This work is a part of NCN Grant No. 2012/05/E/ST2/02352.
WL is supported by the National Centre for Research and Development (Chist-Era Project QUASAR).
The contribution of MM is supported within the International PhD Project ``Physics of future quantum-based information technologies'' grant MPD/2009-3/4 from Foundation for Polish Science. 
TP is supported by the National Research Foundation, the Ministry of Education of Singapore, and the start-up grant of Nanyang Technological University.
RW is supported by the Foundation for Polish Science TEAM project cofinanced by the EU European Regional Development Fund.

\end{document}